\documentstyle[12pt]{article}

\input epsf
\ifx\epsffile\undefined
\newlength{\epsfysize}
\def\epsffile#1#2#3#4]#5{}
\fi

\renewcommand{\thefootnote}{\fnsymbol{footnote}}
\newcommand{\DR}{{\footnotesize{$\overline{{\rm DR}}$}} }
\newcommand{\MS}{{\footnotesize{$\overline{{\rm MS}}$}} }
\def\gtrsim{\raise.3ex\hbox{$>$\kern-.75em\lower1ex\hbox{$\sim$}}}
\def\lesssim{\raise.3ex\hbox{$<$\kern-.75em\lower1ex\hbox{$\sim$}}}

\begin{document}

\begin{flushright}
{\small
SLAC--PUB--7347\\
JHU--TIPAC--96024\\
November 1996\\}
\end{flushright}

\vspace{2cm}

\centerline{GAUGE AND YUKAWA UNIFICATION IN MODELS}
\baselineskip=15pt
\centerline{WITH GAUGE-MEDIATED SUPERSYMMETRY BREAKING}

\baselineskip=32pt

\centerline{\footnotesize JONATHAN A. BAGGER,$^{a}$}
\baselineskip=13pt
\centerline{\footnotesize KONSTANTIN T. MATCHEV,$^{a}$}
\centerline{\footnotesize DAMIEN M. PIERCE\,$^{b}$ and}
\centerline{\footnotesize REN-JIE ZHANG\,$^{a}$}

\baselineskip=22pt

\centerline{\footnotesize\it $^{a}$\,Department of Physics and Astronomy}
\baselineskip=13pt
\centerline{\footnotesize\it Johns Hopkins University}
\centerline{\footnotesize\it Baltimore, Maryland 21218}

\baselineskip=22pt

\centerline{\footnotesize\it $^{b}$\,Stanford Linear Accelerator Center}
\baselineskip=13pt
\centerline{\footnotesize\it Stanford University}
\centerline{\footnotesize\it Stanford, California 94309}

\vspace{1cm}

\abstract{We examine gauge and Yukawa coupling unification in models with
gauge-mediated supersymmetry breaking.  We work consistently to
two-loop order, and include all weak, messenger and unification-scale
threshold corrections.  We find that successful unification requires
unification-scale threshold corrections that are in conflict with the
minimal SU(5) model, but are consistent with the modified missing
doublet SU(5) model for small $\tan\beta$, and large $\tan\beta$ with
$\mu > 0$.}

\vspace{2cm}

\vfill

{\noindent\em Work supported by Department of Energy contract
DE--AC03--76SF00515 and by the U.S. National Science Foundation,
grant NSF-PHY-9404057.}

\pagebreak

\normalsize\baselineskip=15pt
\setcounter{footnote}{0}
\renewcommand{\thefootnote}{\arabic{footnote}}

The apparent unification of the gauge couplings in the minimal
supersymmetric standard model (MSSM) \cite{gauge} has sparked much
interest in supersymmetric extensions to the standard model.  In their
present form, most phenomenologically viable models have two sectors:
a hidden sector, in which supersymmetry is broken, and a visible
sector, which contains the standard-model particles and their
supersymmetric partners.  Supersymmetry breaking is transmitted to the
visible sector by gravitational interactions (as in
supergravity-inspired models) or by standard-model gauge interactions
(as in models with gauge-mediated dynamical supersymmetry breaking).

Models with gauge-mediated supersymmetry breaking are usually
constructed to preserve gauge coupling unification to one-loop order.
In this letter we will report on a closer look at unification in
gauge-mediated models.  We will present the results of a complete
two-loop analysis for gauge and Yukawa coupling unification.  Our
computation takes all one-loop thresholds into account, including
those at the weak, messenger and unification scales.  The thresholds
include finite terms which turn out to be very important for our
precision analysis.

We will present our results in terms of the model-independent
unification-scale threshold corrections $\epsilon_g$ and $\epsilon_b$
\cite{precise}.  These parameters describe conditions that must be
satisfied by any viable unification model.  We will illustrate the
range of these parameters for the minimal \cite{SU5} and (modified)
missing-doublet \cite{MD,MMD} SU(5) models.  We will see that present
precision measurements exclude the minimal model, but are consistent
with gauge and Yukawa unification in the modified missing-doublet
case.

In the simplest models of gauge-mediated supersymmetry breaking
\cite{DN}, the messenger sector contains a set of vector-like fields
which couple only to a standard-model singlet spurion through
trilinear terms in the superpotential.  The vector-like messenger
fields are chosen to transform in $5 + \overline {5}$ or $10 +
\overline {10}$ representations of SU(5).  Requiring the gauge
couplings to remain perturbative restricts attention to at most four
$5+\overline{5}$ or one $10+\overline{10}$ plus one $5+\overline{5}$
pair of fields.  (An additional $5 + \overline{5}$ pair can be
accommodated if the messenger particles are sufficiently heavy.)

We assume that the lowest ($S$) and highest ($F$) components of the
spurion acquire vevs through their interactions with the hidden
sector.  These interactions remove the mass degeneracy of the
messenger superfields and transmit supersymmetry breaking from the
hidden to the visible sector through loop diagrams which contain
spurion insertions.  At the messenger scale, gaugino and soft scalar
masses are induced by one-loop and two-loop diagrams, respectively.
The flavor-blind nature of the gauge interactions ensures that
flavor-changing neutral currents are suppressed.  To this order, the
soft supersymmetry-breaking $A$-parameter is not generated.

The supersymmetric Higgs mass parameter $\mu$ and the soft
supersymmetry-breaking $B$-parameter violate a Peccei-Quinn symmetry
and cannot be generated by standard-model gauge interactions.  We will
assume that they are generated by some minimal mechanism. The region
where $B=0$ is theoretically appealing \cite{B=0} because it gives
rise to a large ratio of vevs ($\tan\beta$) without fine tuning.  In
this region, all CP-violating phases are generated only radiatively,
so CP violation is naturally small.

Our approach is as follows.  We start with the Fermi constant, $G_F$,
the electromagnetic coupling, $\alpha_{em}$, the $Z$-boson mass,
$M_Z$, the \MS strong coupling constant, $\alpha_s(M_Z)$, and the
top-, bottom-quark and tau-lepton pole masses, $m_t$, $m_b$ and
$m_\tau$ (for details, see \cite{BMPZ}).  We then assume a
supersymmetric spectrum and use the full one-loop corrections to
calculate the \DR couplings $g_1$, $g_2$, $g_3$, $\lambda_t$,
$\lambda_b$ and $\lambda_\tau$ for a given value of $\tan\beta$.  We
run these couplings to the messenger scale, $M$, using the two-loop
MSSM renormalization group equations.  At $M$ we fix the gaugino and
soft scalar masses \cite{GMSB}.  We then run the soft parameters back
to the squark mass scale, where we impose electroweak symmetry
breaking and calculate the supersymmetric spectrum.  We iterate the
procedure several times to achieve a consistent solution.

Our calculations of the one-loop threshold corrections include the
finite and logarithmic terms.  The finite corrections, which are often
neglected in the literature, allow a precise determination of the
gauge couplings $g_1$ and $g_2$ at the scale $M_Z$ \cite{precise,precise2}.
The finite corrections to the bottom and tau Yukawa couplings also
play an important role in our analysis.

Once we determine the gauge and Yukawa couplings at the messenger
scale, we extrapolate them to the unification scale, $M_{GUT}$, which
we define to be the scale where $g_1$ and $g_2$ meet.  We use the
usual two-loop beta functions to compute the evolution of the gauge
and Yukawa couplings.  We also include the messenger contributions,
those listed in Ref.~\cite{Hemp,C&M}, and \cite{Blok}
\begin{equation}
\mu {dg_i\over d\mu} = - {g^3_i\over (16 \pi^2)^2}
\biggl( \sum_f D_{if} y_f^2 \biggr) + \ldots
\label{dg}
\end{equation}
\begin{equation}
\mu {d\lambda_a\over d\mu} = (n_5+3n_{10}){\lambda_a\over (16
\pi^2)^2} \biggl( \sum_{i=1}^3 C_{ai} g_i^4 \biggr) + \ldots
\end{equation}
The sum over $f$ runs over {\it all} messenger multiplets, $n_5$
and $n_{10}$ are the number of $5 + \overline{5}$ and $10 +
\overline{10}$ messenger fields, and
\begin{equation}
D_{if} = \left[\begin{array}{ccccc}
{4\over 5} & {6\over 5} & {2 \over 5}&{16\over 5} & {12\over 5}\\[1mm]
0 & 2 & 6 & 0 & 0 \\
2 & 0 & 4 & 2 & 0 \\
\end{array}\right], \qquad f = d,\ \ell,\ q,\ u,\ e,
\end{equation}
\begin{equation}
C_{ai} =
\left[\begin{array}{ccc}
{13\over 15} & 3 & {16 \over 3} \\[1mm]
{7 \over 15} & 3 & {16 \over 3} \\[1mm]
{9 \over 5}  & 3 & 0            \\
\end{array}\right],
\qquad a = t, b, \tau\ .
\end{equation}
At $M_{GUT}$ we set the messenger Yukawas to a common value, $y_m$.
We run the messenger Yukawas back to the messenger scale according to
their one-loop evolution equations,
\begin{equation}
\mu{dy_f\over d\mu} = {y_f\over 16\pi^2}\biggl( 2 y^2_f + T -
4\sum_{i=1}^3 g^2_i C_i(f)\biggr)~,
\label{dy}
\end{equation}
where $T = n_5 (3 y^2_d + 2 y^2_\ell) + n_{10} ( 6 y^2_q + 3 y^2_u +
y^2_e)$ and the $C_i$'s are the quadratic Casimirs, $3Y^2/5,\ 3/4,$
and 4/3 for fundamental representations.  (The messenger-Yukawa
evolution equations can receive additional model-dependent
contributions from the hidden-sector particles.  The extra terms do
not affect the messenger mass splittings, so we can ignore them in our
analysis.  Note that the one-loop equations suffice because the
messenger-sector Yukawas enter our calculation only through the
messenger threshold corrections.)

{}From the set of $y_f(M)$, we determine the messenger-particle mass
spectrum and compute the messenger-scale threshold corrections to the
gauge couplings,
\begin{equation}
\Delta \alpha_i^{-1}(M) =
\sum_f {D_{if}\over 8\pi} \biggl[\ln {M_f^2\over M^2}
+{1\over 6}\ln\biggl(1-{\Lambda^2\over M_f^2}\biggr)\biggr]~,
\end{equation}
where $\Lambda = F/S$ is the supersymmetry-breaking scale and $M_f$ is
the messenger fermion mass.  The second term in the brackets is small
for $\Lambda/M_f \ll 1$, in which case there is a near degeneracy
among the masses in the vector-like supermultiplets.  Note that there
are no messenger-scale Yukawa thresholds to this order.

We iterate this procedure to find a consistent solution in the region
between $M$ and $M_{GUT}$.  At $M_{GUT}$ we define the threshold
corrections for the gauge and Yukawa couplings, $\epsilon_g$ and
$\epsilon_b$, as follows,
\begin{eqnarray}
g_3(M_{GUT}) & = & g_1(M_{GUT}) (1 + \epsilon_g), \nonumber \\ [1mm]
\lambda_b(M_{GUT}) & = & \lambda_\tau(M_{GUT}) (1 + \epsilon_b)~.
\end{eqnarray}
The parameters $\epsilon_g$ and $\epsilon_b$ describe the
unification-scale threshold corrections that are necessary to achieve
unification in any particular model.  In what follows, we will
indicate the allowed ranges of $\epsilon_g$ and $\epsilon_b$ for two
of the simplest unification models, the minimal and the modified
missing-doublet SU(5) models.

In the minimal SU(5) model, the unification-scale gauge threshold
correction is \cite{precise,SU5 threshold},
\begin{equation}
\epsilon_g = {3 g^2_{GUT}\over 40 \pi^2} \ln \biggl({M_H\over M_{GUT}}
\biggr)~,
\end{equation}
where $M_H$ is the mass of the color-triplet Higgs multiplet that
mediates nucleon decay.  Generally, $M_H$ is bounded from below by the
proton decay limits \cite{nucleon}, which imply $M_H \gtrsim M_{GUT}$,
so $\epsilon_g \gtrsim 0$.

The missing-doublet model is an alternative SU(5) theory in which the
heavy color-triplet Higgs particles are split naturally from the light
Higgs doublets \cite{MD}.  This requires large SU(5) representations,
such as the $75$ and $50 + \overline{50}$, so the SU(5) coupling $g_5$
diverges below the Planck scale.  The modified missing-doublet (MMD)
model solves this problem for $n_5\le1$ by lifting the mass of the $50
+\overline{50}$ to the Planck scale and suppressing the nucleon decay
rate through an extra Peccei-Quinn symmetry \cite{MMD}.  In this way
the modified missing doublet model can accommodate two color-triplet
Higgs particles with masses between $10^{13} - 10^{15}$ GeV.

In the modified missing-doublet model, the unification-scale gauge
threshold can be written as \cite{precise,MMD,threshold}
\begin{equation}
\epsilon_g = {3 g^2_{GUT}\over 40\pi^2} \Biggl\{ \ln\biggl( {M_H^{\rm
eff} \over M_{GUT}}\biggr)- 9.72\Biggr\}~,
\end{equation}
where $M_H^{\rm eff}$ is the effective mass that enters the proton
decay amplitude, so the previous lower bounds on $M_H$ apply here as
well.  In the MMD case, the effective mass is also bounded from above,
$M_H^{\rm eff} \lesssim 10^{20}$ GeV \cite{MMD}.

The Yukawa threshold in minimal SU(5) can be written as follows
\cite{precise,Yukawa},
\begin{equation}
\epsilon_b = {1\over 16\pi^2}\Biggl\{
 4 g^2_{GUT} \biggl[\ln\biggl({M_V\over M_{GUT}}\biggr)
- {1\over 2}\biggr]\ -\ \lambda_t^2(M_{GUT})
\biggl[\ln\biggl({M_H\over M_{GUT}}\biggr)-{1\over 2}\biggr]\Biggr\}~,
\label{yth}
\end{equation}
where $M_V$ is the mass of a superheavy SU(5) gauge boson.  For the
minimal SU(5) model, the most stringent lower limit on $M_V$ comes
from requiring that the $5+\overline{5}$ Higgs coupling remain
perturbative to the Planck scale.  This implies $M_V \gtrsim 0.5 M_H$
\cite{nucleon}.  We take the upper limit on $M_V$ to be the Planck
scale, $M_V \le 10^{19}$ GeV.

For the modified missing-doublet model, the Yukawa threshold has the
same form as eq.~(\ref{yth}), with the color-triplet Higgs mass,
$M_H$, replaced by the effective mass, $M_H^{\rm eff}$.  In this case,
the lower limit on $M_V$ comes from proton decay experiments, which
imply $M_V/g_{\rm GUT} \gtrsim 3.8\times10^{15}$ GeV \cite{XYboson}.
As before, we impose $M_V\le10^{19}$ GeV. Hence, both models have the
same upper limit on $\epsilon_b$, but the lower limit in the MMD model
is lower, by virtue of the fact that $M_V$ can be smaller and $M_H$
larger.

In what follows, we present our results for gauge-mediated models.  In
particular, we calculate $\epsilon_g$, $\epsilon_b$, $\alpha_{GUT}$
and $M_{GUT}$ as functions of the input parameters, which we take to
be $\tan\beta$, the numbers $n_5$ and $n_{10}$, the
supersymmetry-breaking scale $\Lambda$, the messenger scale $M$, and
the messenger Yukawa at the unification scale, $y_m$.  To examine
bottom-tau unification, we fix the sign of $\mu$ to be positive.

We find the range of $\alpha_{GUT}$ and $M_{GUT}$ by scanning over the
parameter space, with $m_t = 175$ GeV, $m_b = 4.9$ GeV, $\Lambda \le
300$ GeV, $1.03 \le M/\Lambda \le 10^4$, $0.03 \le y_m \le 3.0$ and
$\tan\beta$ in the allowed range.  For the case $n_5=1$, we determine
$\alpha_{GUT} \simeq (0.044 - 0.054)$ and $M_{GUT} \simeq (1.5 - 5.0)
\times 10^{16}$ GeV.  For $n_5 = n_{10} = 1$, we find $\alpha_{GUT}
\simeq (0.062 - 0.28)$ and $M_{GUT} \simeq (1.2 - 7.0) \times 10^{16}$
GeV.

\begin{figure}[t]
\epsfysize=3in
\epsffile[-40 225 100 535]{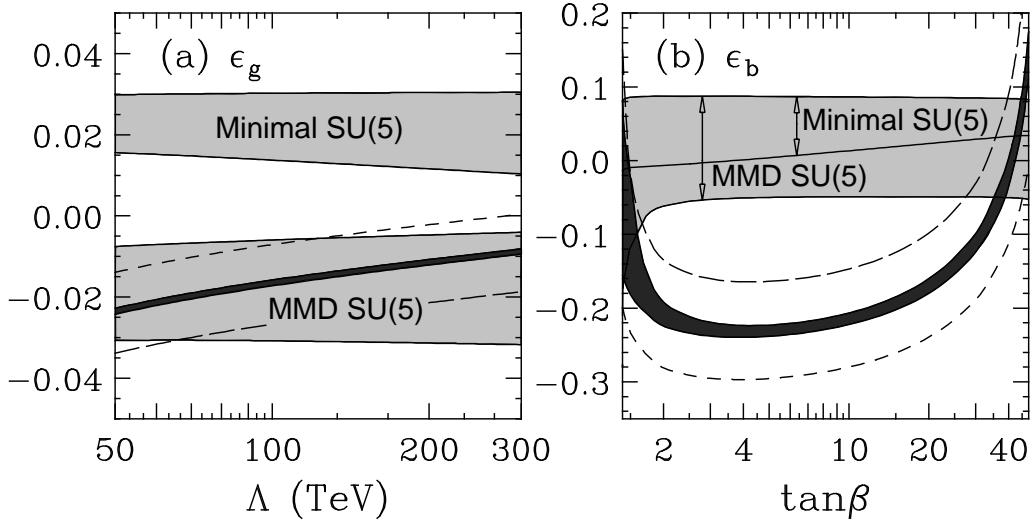}
\begin{center}
\parbox{5.5in}{
\caption[]{The unification-scale threshold corrections with $n_5=1,
{}~\mu>0,\ M/\Lambda=2,$ and $y_m=1$. (a) The gauge coupling
unification-scale threshold correction $\epsilon_g$ versus $\Lambda$,
for $\tan\beta=20$, and $\alpha_s(M_Z)= 0.118$ (black band), 0.124
(short-dashed) and 0.112 (long-dashed).  (b) The Yukawa coupling
unification-scale threshold correction, $\epsilon_b$, versus
$\tan\beta$, for $\Lambda=100$ TeV and the same values for
$\alpha_s(M_Z)$ as in (a).  In each case, the black band is obtained
by varying the top mass from 170 to 180 GeV.  The shaded regions
indicate the allowed range for (a) $\epsilon_g$ and (b) $\epsilon_b$
in the minimal and modified missing-doublet SU(5) models.
\label{f1}}}
\end{center}
\end{figure}

In Fig.~\ref{f1} we plot $\epsilon_g$ and $\epsilon_b$ for $n_5=1$,
$M/\Lambda=2$, $m_b = 4.9$ GeV and $y_m=1$, versus $\Lambda$ and
$\tan\beta$, respectively.  In (a) we choose $\tan\beta = 20$, while
in (b) we take $\Lambda = 100$ TeV.  In each case the short-dashed
(long-dashed) lines correspond to $\alpha_s(M_Z)=0.124$ (0.112).  The
black bands correspond to $\alpha_s(M_Z) = 0.118$ with $m_t$ varying
from $170$ to $180$ GeV.  The uncertainty in $\epsilon_b$ from varying
$m_b = 4.9 \pm 0.3$ GeV is almost the same as that from changing
$\alpha_s(M_Z) = 0.118 \pm 0.006$.

In Fig.~\ref{f1}(a) we also show the allowed values for $\epsilon_g$
in the minimal and modified missing-doublet SU(5) models.  The region
of allowed $\epsilon_g$ in the modified missing-doublet model almost
completely overlaps the region with $\alpha_s(M_Z) = 0.118 \pm 0.006$.
In contrast, we see that minimal SU(5) is inconsistent with
$\alpha_s(M_Z)$ by more than 2$\sigma$.

For $n_5=1$ we find that the messenger sector corrections decrease
$\epsilon_g$.  The change is induced by the messenger thresholds and
the differences in the two-loop gauge coupling evolution.  Both of
these effects are of approximately equal importance.

{}From Fig.~\ref{f1}(a) we see that raising the supersymmetry-breaking
scale $\Lambda$ decreases the size of the gauge-coupling
unification-scale threshold.  This is because the superpartner masses
scale with $\Lambda$, and larger masses decrease the size of the
required thresholds \cite{precise,precise2}.

Figure \ref{f1}(b) illustrates the well-known fact that bottom-tau
unification can only be achieved for very small ($ \lesssim 1.8$) or
rather large ($\gtrsim 35$) $\tan\beta$.  (Very large values of
$\tan\beta$ are excluded by the requirement of proper electroweak
symmetry breaking.)  Figure \ref{f1}(b) also shows the allowed bands for
$\epsilon_b$ in the minimal and modified missing-doublet SU(5) models.

As above, we can compare this plot to the case with no messengers.
There, one typically finds that the bottom and tau Yukawa couplings
meet much earlier than the scale $M_{GUT}$, so a rather large and
negative threshold correction $\epsilon_b$ is required.  For the case
at hand, the extra messenger multiplets change the Yukawa evolution
equations at two loops.  More importantly, however, they also increase
the gauge couplings, which feed into the Yukawa evolution equations
and cause the bottom and tau couplings to meet even earlier.  This
makes $\epsilon_b$ even more negative.

Fortunately, at large $\tan\beta$ there are significant {\em finite}
threshold corrections to the bottom (and to a smaller extent, tau)
Yukawa couplings \cite{BMPZ}.  These corrections, which are
proportional to $\mu \tan\beta$, are sufficiently important to permit
bottom-tau unification at large $\tan\beta$ for $\mu > 0$. (The case
$\mu < 0$ is completely excluded at large $\tan\beta$.)  These finite
corrections were omitted in the analysis of Ref.~\cite{C&M}, which
came to a different conclusion.

For $n_5 = 1$, the value of $\epsilon_g$ is not significantly affected
by changes in $\tan\beta$ or $M/\Lambda$.  At the smallest values of
$\tan\beta$, $\epsilon_g$ increases by about 0.5\%, while for
$M/\Lambda=10^4$, $\epsilon_g$ increases by about 0.2\%.  The
parameter $\epsilon_b$ is more sensitive to changes in $M/\Lambda$.
For $M/\Lambda = 10^4$, the $\epsilon_b$ curve is 2.5 to 3\% higher at
intermediate $\tan\beta$, and rises to $+20\%$ at $\tan\beta \simeq
40$.

\begin{figure}[t]
\epsfysize=3in
\epsffile[-30 225 110 535]{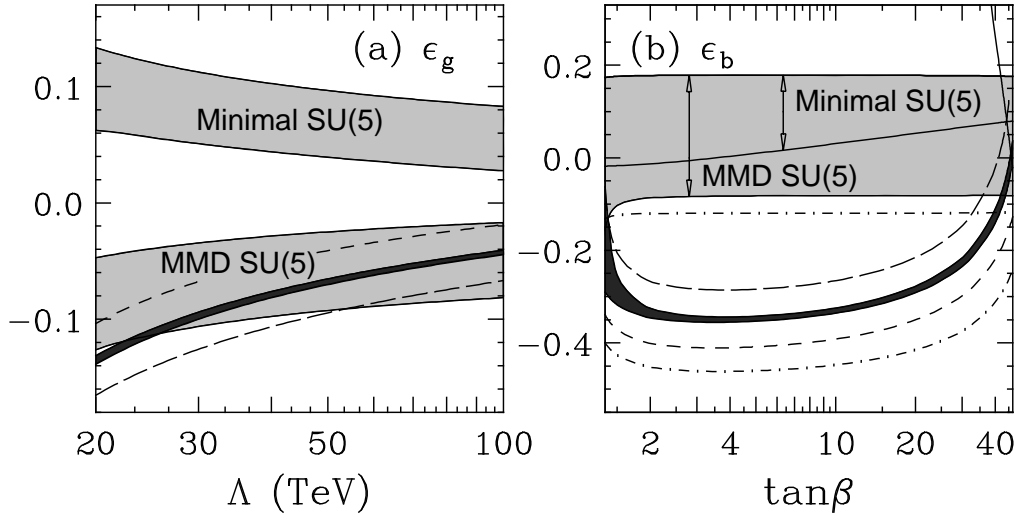}
\begin{center}
\parbox{5.5in}{
\caption[]{The same as Fig.~\ref{f1}, except that $n_5=n_{10}=1$, and
in (b) $\Lambda=$50 TeV and $M/\Lambda=100$.  The dot-dashed lines
indicate $\epsilon_b$ and its lower limit in the MMD model for
$M/\Lambda = 2$.  The line in the upper right-hand corner of (b)
describes the top-quark threshold, $\epsilon_t$, for $M/\Lambda =
100$.
\label{f2}}}
\end{center}
\end{figure}

In Fig.~\ref{f2} we plot $\epsilon_g$ and $\epsilon_b$ for the case of
$n_5 = n_{10} = 1$, versus $\Lambda$ and $\tan\beta$ respectively.
The other parameters are fixed as in Fig.~\ref{f1}, except that in
Fig.~\ref{f2}(b), $\Lambda = 50$ TeV (to keep the scalar masses
unchanged) and $M/\Lambda = 100$. (Two $M/\Lambda=2$ curves are shown
in dotted lines.)  Figure \ref{f2}(a) shows that everything shifts
because of the larger $\alpha_{GUT}$, but the overlap between the band from
the MMD model and the allowed region for $\alpha_s(M_Z)$ is still
almost complete.  In this case, increasing $M/\Lambda$ to $10^4$
significantly changes Fig.~2(a).  The central value of $\epsilon_g$
runs from $-4\%$ for $\Lambda = 20$ TeV to $-1.5\%$ for $\Lambda =
100$ TeV.  The band for the MMD model is such that the required value
of $\epsilon_g$ lies entirely within the band.  (Note, however, that
$n_5 = n_{10} = 1$ gives rise to nonperturbative couplings above
$M_{GUT}$ in the MMD case.)

The change in Fig.~\ref{f2}(b) as compared to Fig.~\ref{f1}(b) is more
dramatic.  Because the gauge couplings are even larger than in the
previous case, bottom-tau unification turns out to be barely possible
for $M/\Lambda = 2$ (dot-dashed lines).  Note, however, that there is
still a significant region for unification in the missing doublet
model with $M/\Lambda = 100$.  In Fig.~\ref{f2}(b) we also show the
necessary threshold, $\epsilon_t$, for top-tau Yukawa unification.  We
see from Fig.~\ref{f2}(b) that the top, bottom and tau couplings unify at
the largest values of $\tan\beta$ (in the region where $B \simeq 0$).
Such a unification is expected in SO(10) models.  However, the
thresholds in any particular SO(10) model must be calculated to be
sure the model is consistent with data.

\end{document}